\documentclass{article}
\usepackage{spconf,amsmath,graphicx}
\usepackage{amssymb,amsthm,amsfonts}
\usepackage{nicefrac}
\usepackage{hyperref}
\usepackage[table]{xcolor}
\usepackage[export]{adjustbox}
\usepackage{booktabs}
\usepackage{flushend} 

\DeclareMathOperator*{\argmax}{arg\,max}
\DeclareMathOperator*{\argmin}{arg\,min}
\newcommand\norm[1]{\left\lVert#1\right\rVert}

\usepackage{fancyhdr}
\fancypagestyle{firstpage}{
  \fancyhf{}
  \fancyhead[C]{\vspace{-2.2cm} \small Accepted for presentation at the IEEE Int. Conf. on Image Processing (ICIP), Anchorage, AK, USA, September 19-22, 2021.\vspace{0.2cm}\\© 2021 IEEE. Personal use of this material is permitted. Permission from IEEE must be obtained for all other uses, in any current or future media, including reprinting/republishing this material for advertising or promotional purposes, creating new collective works, for resale or redistribution to servers or lists, or reuse of any copyrighted component of this work in other works. \vspace{0.3cm}} 
}

\title{CALTeC: Content-Adaptive Linear Tensor Completion for Collaborative Intelligence}
\name{Ashiv Dhondea, Robert A. Cohen, and Ivan V. Baji{\'c}}
\address{School of Engineering Science, Simon Fraser University, Burnaby, BC, Canada}
\begin{document}
%
\maketitle
\begin{abstract}
In collaborative intelligence, an artificial intelligence (AI) model is typically split between an edge device and the cloud. Feature tensors produced by the edge sub-model are sent to the cloud via an imperfect communication channel. At the cloud side, parts of the feature tensor may be missing due to packet loss. In this paper we propose a method called Content-Adaptive Linear Tensor Completion (CALTeC) to recover the missing feature data. The proposed method is fast, data-adaptive, does not require pre-training, and produces better results than existing methods for tensor data recovery in collaborative intelligence.
\end{abstract}

\thispagestyle{firstpage}

\section{Introduction}
\label{sec:intro}
Collaborative intelligence (CI)~\cite{neurosurgeon, jointdnn} has emerged as a promising way to bring ``AI to the edge.'' In CI, typically, a Deep Neural Network (DNN) is split between an edge device and the cloud. The input signal (image, video, ...) is processed by the edge sub-model, which usually consists of the initial layers (front-end) of a DNN. The resulting feature tensor is transmitted to the cloud, where the remaining DNN layers (back-end) complete the analysis task. Such an approach has shown potential for latency and energy savings compared to purely cloud-based or edge-based analysis solutions~\cite{neurosurgeon, jointdnn}.

To make efficient use of the communication channel in CI, the features produced by the edge sub-model should be compressed. The topic of intermediate feature compression has attracted considerable interest in recent years~\cite{choi2018deep,eshratifar2019bottlenet,Chen20,Duan2020VideoCF,cohen2020lightweight}, and is being explored in the standardization community under the title ``Video Coding for Machines'' (VCM)~\cite{vcm_call_for_evidence}, as well as JPEG-AI~\cite{JPEG-AI_use_cases}. 

The focus of this work, however, is on error/loss resilience in CI. The communication channel across which the intermediate feature tensor is transmitted is imperfect. This will result in bit errors at the physical layer, which will turn into packet loss at the transport/application layer. Missing feature data will affect the accuracy of the analysis performed by the back-end. The topic of error/loss resilience in CI is still largely unexplored. Among the existing work on the topic,~\cite{choi_neural_2019,BottleNet++} studied bit error resilience at the physical layer via joint source-channel coding of intermediate features. In~\cite{unnibhavi2018dfts}, packet-based feature transmission was considered, and several simple packet loss recovery schemes, such as nearest-neighbor and bilinear interpolation, were explored.   

In~\cite{Bragile2020}, more advanced methods for missing feature data recovery were explored. Specifically, two generic tensor completion methods called Simple Low Rank Tensor Completion (SiLRTC) and High accuracy Low Rank Tensor Completion (HaLRTC)~\cite{liu2012tensor} were adapted to recover missing packets in a deep feature tensor. In addition, a method called Adaptive Linear Tensor Completion (ALTeC)~\cite{Bragile2020} was proposed, and shown to be much faster and at least as accurate as SiLRTC and HaLRTC. However, one issue with ALTeC is that it needs to be pre-trained for a specific intermediate feature tensor of a specific DNN model. Applying it to another layer or another DNN model requires re-training. 

In this paper, we propose a new method called Content Adaptive Linear Tensor Completion (CALTeC) for the recovery of missing data in deep feature tensors. CALTeC manifests both the generic nature of SiLRTC/HaLRTC and the speed characteristics of ALTeC. CALTeC is based on estimating a linear (or, more precisely, an affine) relationship between the missing feature packet and an available, spatially collocated, feature packet in the most similar channel of the feature tensor. The estimated relationship enables approximate recovery of the missing feature packet. 

The paper is organized as follows. Section~\ref{sec:pktquant} establishes the notation and gives the background information needed for the remainder of the paper. The proposed method is presented in Section~\ref{sec:tc}. Experiments are described in Section~\ref{sec:experiments}, followed by conclusions in Section~\ref{sec:conclusions}. 
An implementation of the proposed method is also provided.\footnote{ \url{https://github.com/AshivDhondea/DFTS_TF2}}

\section{Preliminaries}
\label{sec:pktquant}
\begin{figure}
\centering
\includegraphics[width=7.4cm]{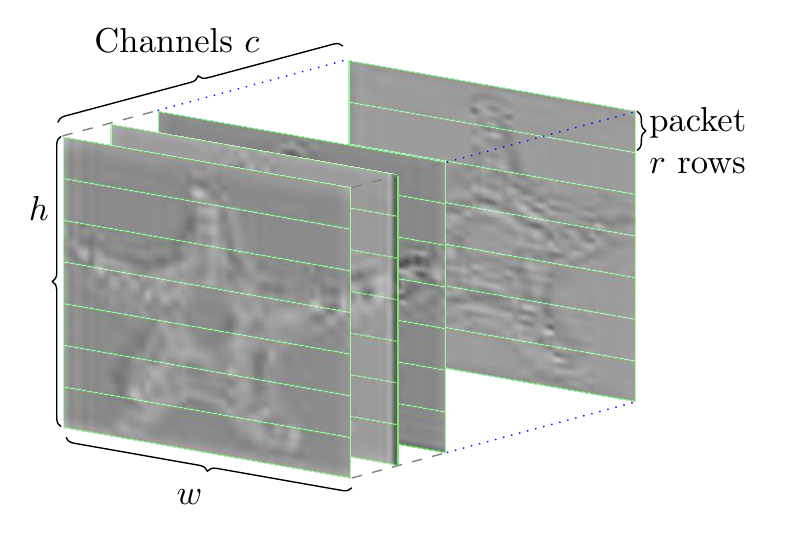}
    \caption{A tensor from layer \texttt{add\_1} of ResNet-18. Several consecutive rows in each channel form a feature data packet.}
    \label{fig:tensorviz}
\end{figure}

A feature tensor with height $h$, width $w$, and the number of channels $c$ will be denoted $\mathcal{X} \in \mathbb{R}^{h \times w \times c}$. An illustration of a feature tensor extracted from layer \texttt{add\_1}
of ResNet\nobreakdash-18~\cite{ResNet} is shown in Fig.~\ref{fig:tensorviz}.
Layer \texttt{add\_1} corresponds to the output of the shortcut connection and element-wise addition applied to the final output of the \texttt{conv2\_x} layers as denoted in~\cite[Table~1]{ResNet}.
Feature data packets are formed by partitioning each tensor channel into groups of $r$ rows, as indicated by green lines in the figure. Such packetization is inspired by common practice in video streaming where video frames are encoded into packets in rows of macroblocks~\cite{video2002}. In our experiments we used $r\in\{4,8\}$. If the number of rows ($h$) in a channel is not divisible by $r$, the last packet is zero-padded to $r$ rows.

The $i$-th packet in the $j$-th tensor channel is denoted $\mathbf{X}_i^{(j)} \in \mathbb{R}^{r \times w}$, for $j \in \left\{0, 1,  \dotsc, c-1\right\}$ 
The vectorized version of the packet is denoted $\mathbf{x}_i^{(j)} = \text{Vec}\left(\mathbf{X}_i^{(j)}\right) \in \mathbb{R}^{(r\cdot w) \times 1}$. In our simulations, the tensors were min-max quantized~\cite{choi2018deep} to 8 bits/element. Further compression of the data in each packet is possible, but since our focus here is not on compression, no further quantization or entropy coding is used.  


The Gilbert-Elliott (GE) channel model~\cite{5755057} is used to simulate the packet loss. This is a two-state Markov model with a good $(G)$ and a bad $(B)$ state. It can be parameterized by the burst loss probability $P_B$ and the average burst length $L_B$. The conversion from $P_B$ and $L_B$ to state transition probabilities is as follows: $p_{B\to G}=1/L_B$, $p_{G\to B}=P_B/(L_B(1-P_B))$, $p_{B\to B}=1-p_{B\to G}$, and $p_{G\to G} = 1-p_{G\to B}$. 





When a feature tensor is transmitted, some of its packets will be lost, depending on the channel realization. The received packets are used to reconstitute the available part of the tensor. The available data is inverse quantized, and the missing data is recovered using one of the tensor completion methods.
Next, we describe our proposed method for missing data recovery.  
\section{Proposed method}
\label{sec:tc}
Recovering missing values in a corrupted deep feature tensor leads to more tensor data that can be exploited by the back-end sub-model, which generally improves inference accuracy~\cite{Bragile2020}. As seen in Fig.~\ref{fig:tensorviz}, there is a considerable amount of redundancy among tensor channels, which can be used to recover missing data, at least approximately. To recover missing data, the proposed CALTeC proceeds as follows.

Let $\mathbf{x}_i^{(j)}$ be a missing packet in channel $j$, and let $\mathbf{x}_{i'}^{(j)}$ be its spatially-closest correctly received neighboring packet in the same channel. If there are two equally close available neighbors (above and below), the one below is selected. If a missing packet does not have any available spatial neighbors in the same channel, i.e. when the entire tensor channel is lost, then all elements in that channel are simply set to zero. This is a rare event, except for large values of $L_B$ and $P_B$.

The idea behind CALTeC is to find the another channel $k$ in the tensor that is locally similar to channel $j$, and transform its data to recover the missing data in channel $j$.
To measure local similarity, we use the Pearson correlation coefficient. A search is executed over all channels $k$ in which both packets $i$ and $i'$ are available, as 
\begin{equation}
    k^* = \underset{k}{\argmax} \frac{\left(\mathbf{x}_{i'}^{(j)}-\mu_{i'}^{(j)}\right)^\top \left(\mathbf{x}_{i'}^{(k)}-\mu_{i'}^{(k)}\right)}{\sigma_{i'}^{(j)}\sigma_{i'}^{(k)}},
    \label{eq:Pearson_coefficient}
\end{equation}
where $\mu_{i'}^{(j)}$ and $\mu_{i'}^{(k)}$ are the sample means of values in packets $\mathbf{x}_{i'}^{(j)}$ and $\mathbf{x}_{i'}^{(k)}$, respectively, and $\sigma_{i'}^{(j)}$ and $\sigma_{i'}^{(k)}$ are their standard deviations. 

Once the locally most similar channel $k^*$ is found, we compute the coefficients of an affine transformation that approximates $\mathbf{x}_{i'}^{(j)} \approx a\cdot \mathbf{x}_{i'}^{(k^*)} + b$. The coefficents are found by least-squares minimization 
\begin{equation}
    (a^*, b^*) = \underset{(a,b)}{\argmin} \norm{\mathbf{x}_{i'}^{(j)} - \left(a\cdot \mathbf{x}_{i'}^{(k^*)} + b\right)}_2^2.
    \label{eq:affine}
\end{equation}
Equation~(\ref{eq:affine}) has a closed-form solution, which can be easily seen once we re-write the cost function. First, let $\mathbf{a}=[a,b]^\top$, and define $\mathbf{Y}=\left[\mathbf{x}_{i'}^{(k^*)} \, | \, \mathbf{1}\right] \in \mathbb{R}^{(r\cdot w) \times 2}$, where $\mathbf{1} \in \mathbb{R}^{(r\cdot w) \times 1}$ is a column vector of all ones. Then \eqref{eq:affine} becomes
\begin{equation}
    \mathbf{a}^* = \underset{\mathbf{a}}{\argmin} \norm{\mathbf{x}_{i'}^{(j)} - \mathbf{Y}  \mathbf{a}}_2^2,
\end{equation}
which has a well-known solution
\begin{equation}
    \mathbf{a}^* = \left(\mathbf{Y}^\top\mathbf{Y}\right)^{-1}\mathbf{Y}^\top\mathbf{x}_{i'}^{(j)}.
    \label{eq:ls_affine_solution}
\end{equation}

Finally, when the affine parameters $\mathbf{a}^*=[a^*,b^*]^\top$ are computed, the missing packet $\mathbf{x}_i^{(j)}$ is estimated as the affine transformation of the collocated packet in channel $k^*$:
\begin{equation}
    \widehat{\mathbf{x}}_i^{(j)} = a^*\cdot \mathbf{x}_i^{(k^*)} + b^*.
    \label{eq:recovered_packet}
\end{equation}
The above process is repeated for all missing packets in all channels. 

\begin{figure}
    \centering
    \includegraphics[width=8.5cm]{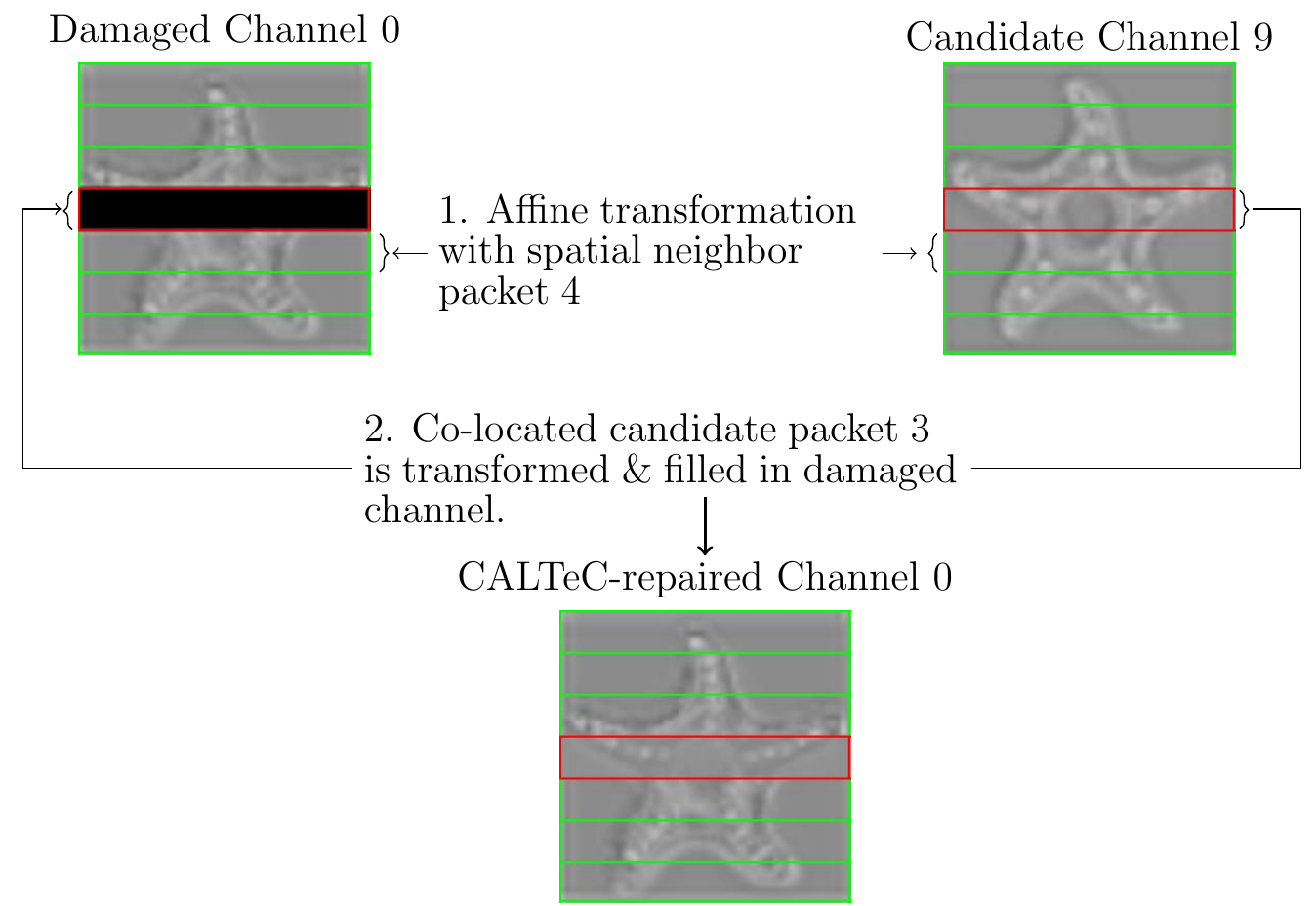}
    \caption{Illustration of CALTeC repairs on corrupted channel 0 of a deep feature tensor.}
    \label{fig:lumimap}
\end{figure}

To illustrate the process, Fig.~\ref{fig:lumimap} shows a missing packet (black rectangle) in channel $j=0$ of a deep feature tensor from layer \texttt{add\_1} of ResNet\nobreakdash-18. Packets are indexed starting from $0$, so the missing packet is $\mathbf{x}_3^{(0)}$. Its neighboring packet, $\mathbf{x}_4^{(0)}$, is used to find the most locally similar channel according to~\eqref{eq:Pearson_coefficient}, which in this example is channel $k^*=9$. Then the affine transformation between $\mathbf{x}_4^{(0)}$ and $\mathbf{x}_4^{(9)}$ is estimated according to~\eqref{eq:affine}-\eqref{eq:ls_affine_solution}. Finally, the missing packet $\mathbf{x}_3^{(0)}$is estimated from $\mathbf{x}_3^{(9)}$ according to~\eqref{eq:recovered_packet} and is used to fill the gap in the damaged channel 0.

\section{Experiments}
\label{sec:experiments}
Experiments were run on the same 882 images used in~\cite{unnibhavi2018dfts}. 
 These images come from 10 classes of the ImageNet~\cite{imagenet2015} test set. 
Experiments were based on a pre-trained ResNet\nobreakdash-18 model.\footnote{ \url{https://github.com/qubvel/classification_models}} Two split points were considered:  \verb|add_1| and \verb|add_3|,
which correspond to the outputs of the shortcut connection and element-wise addition applied to the final output of the \texttt{conv2\_x} and \texttt{conv3\_x} layers, respectively, as denoted in ~\cite[Table~1]{ResNet}.
The feature tensor at the output of \verb|add_1| has dimensions $56 \times 56 \times 64$, and the tensor output by \verb|add_3| has dimensions $28 \times 28 \times 128$. Tensors were quantized to 8 bits per tensor element. To packetize tensor elements, we used $r=8$ rows/packet for the \verb|add_1| tensor and $r=4$ rows/packet for the  \verb|add_3| tensor. In both cases this gave 7 packets per tensor channel. 

The deep feature tensor packet transmission simulator DFTS~\cite{unnibhavi2018dfts} was updated for compatibility with TensorFlow 2 and Python 3.6 in this study. Monte Carlo experiments were run on the Compute Canada cluster \textit{Cedar}. Identical hardware resources were requested in all experiments: 6 CPU cores (Intel E5-2650 v4 Broadwell @ 2.2GHz) and 1 NVIDIA P100 Pascal GPU. The CPU and GPU RAM reserved were, respectively, 32 GB and 12 GB. Speed measurements in Table \ref{tab:methods:speed} were limited to the tensor completion step of a packet transmission experiment. Tensor completion with all four methods is done on the CPU only.

\begin{table}[b]
\caption{Tensor completion methods used in the experiments.}
\vspace{4pt}
    \label{tab:methods}
    \centering
    \begin{tabular}{l|c|c}
         Method & Iterative? & Requires pre-training? \\
         \hline
         SiLRTC~\cite{liu2012tensor}& Yes & No \\
         HaLRTC~\cite{liu2012tensor} & Yes & No \\
         ALTeC~\cite{Bragile2020} & No & Yes \\
         CALTeC & No & No
    \end{tabular}
\end{table}

Transmission of packets was simulated using the GE channel model~\cite{5755057} with burst loss probabilities $P_B \in \{0.01, 0.10, 0.20, 0.30\}$ and average burst lengths of $L_B \in \{1, 2, \dotsc, 7\}$. These can be converted to GE transition probabilities as explained in Section~\ref{sec:pktquant}. In particular, 10 channel realizations across our test set of 882 images were simulated for each pair of $(P_B, L_B)$ values. Each corrupted tensor was saved and then presented to each tensor completion method, so that each method had to deal with the same loss patterns.

Four tensor completion methods were compared: SiLRTC~\cite{liu2012tensor}, HaLRTC~\cite{liu2012tensor}, ALTeC~\cite{Bragile2020}, and the proposed CALTeC. Their main characteristics are shown Table~\ref{tab:methods}. Originally, ALTeC was described for packets consisting of one row of tensor elements per packet~\cite{Bragile2020}.
To run in our simulation scenario, we trained ALTeC for $r\in\{4,8\}$ rows of 8-bit (inverse) quantized tensor data per packet, depending on the split layer.
In the results that follow, `No Loss' is the performance of the system without packet loss or quantization, 
i.e., the performance of the baseline model (ResNet\nobreakdash-18) without splitting into edge and cloud sub-models. `No Completion' means no tensor completion, that is, tensor elements from missing packets are simply set to zero.  

\begin{figure}[t]
	\centering
		\begin{minipage}[b]{0.49\linewidth}
		\centering
		\centerline{\includegraphics[width=4.3cm]{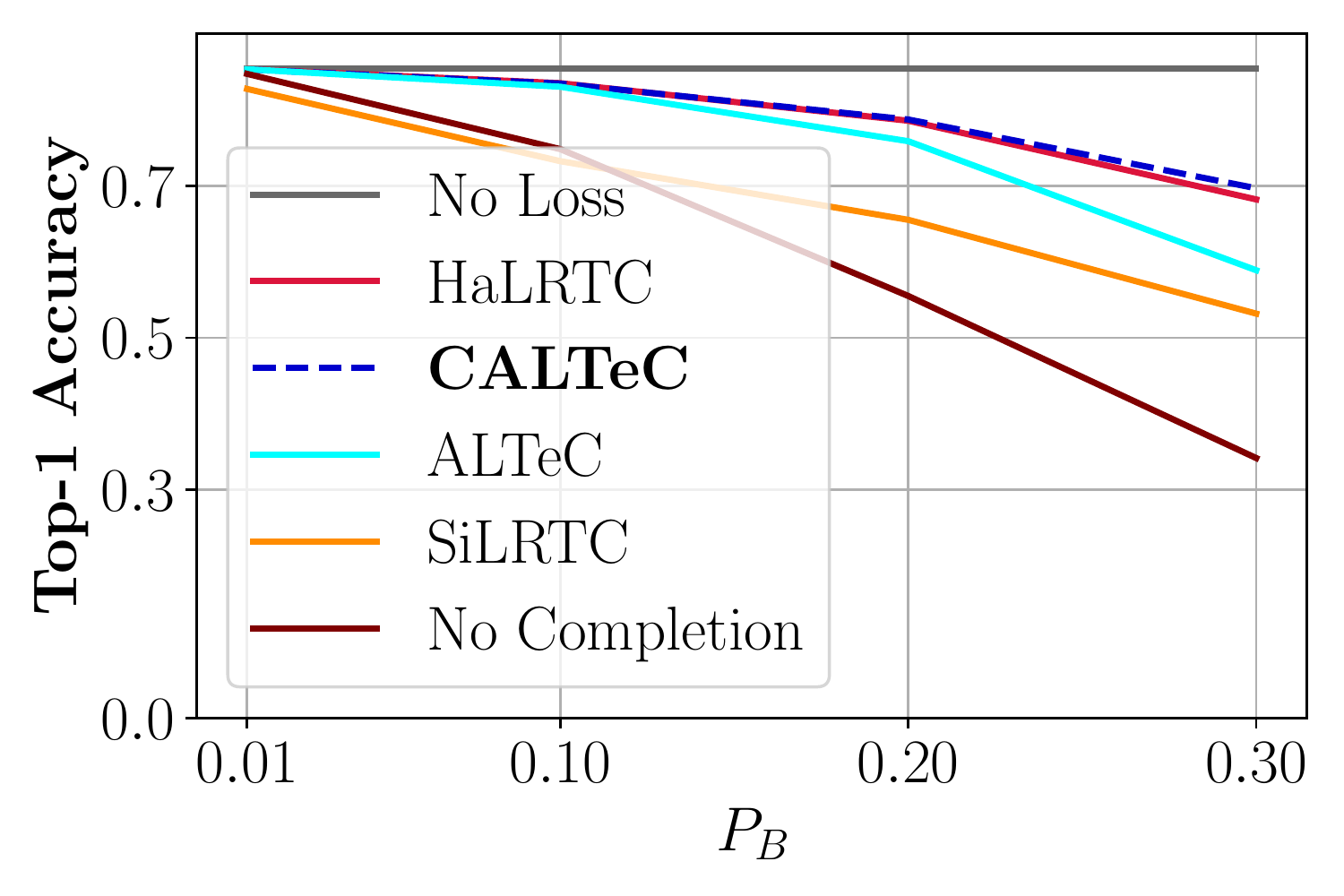}}
		\centerline{(a) Default settings \texttt{add\_1}.}\medskip
	\end{minipage}
\begin{minipage}[b]{0.49\linewidth}
	\centering
	\centerline{\includegraphics[width=4.3cm]{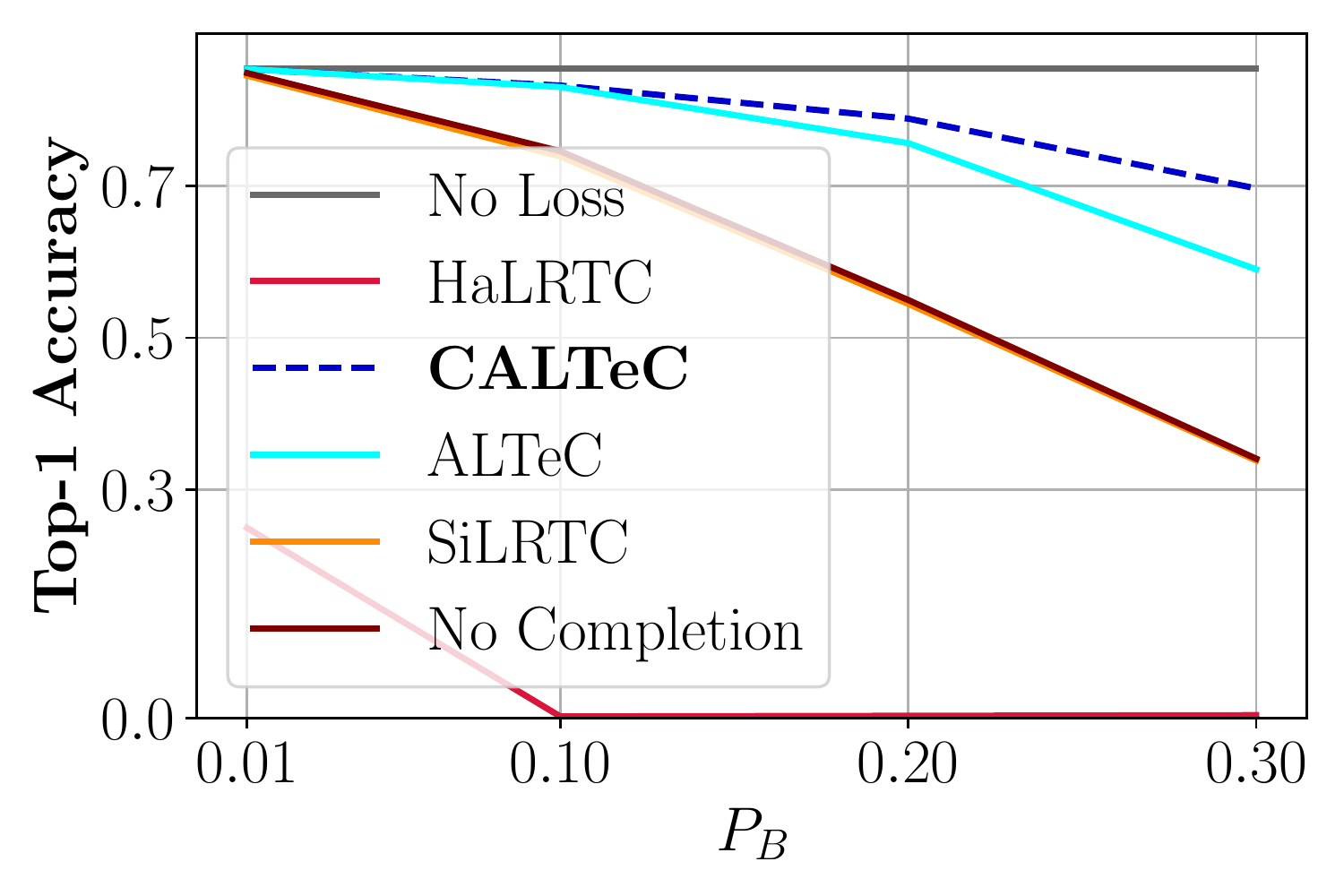}}
	\centerline{(b) Speed-matched \texttt{add\_1}.} \medskip
\end{minipage}
\hfill
	\begin{minipage}[b]{0.49\linewidth}
	\centering
\centerline{\includegraphics[width=4.3cm]{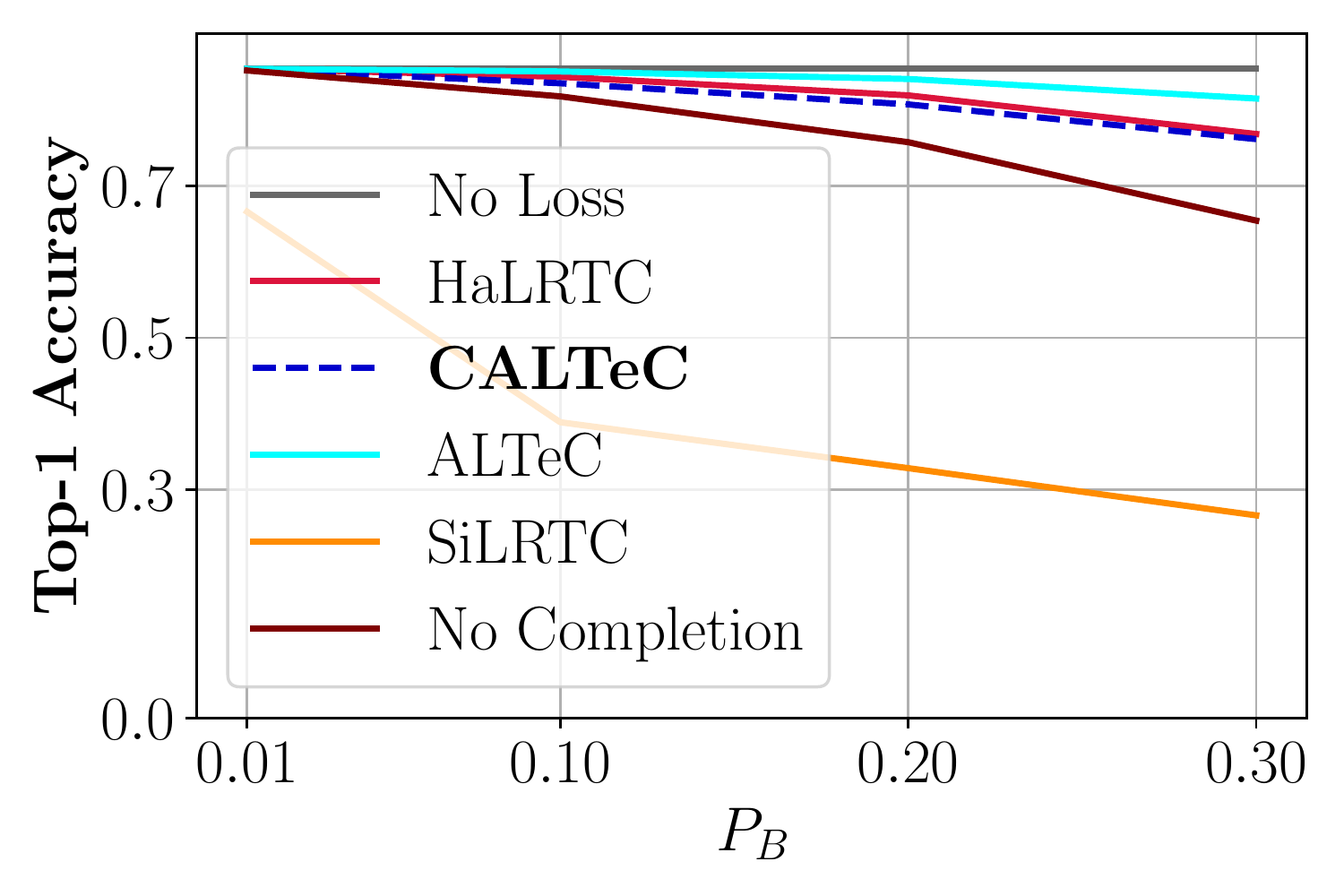}}
	\centerline{(c) Default settings \texttt{add\_3}.}
    \end{minipage}%
	\begin{minipage}[b]{0.49\linewidth}
		\centering
		\centerline{\includegraphics[width=4.3cm]{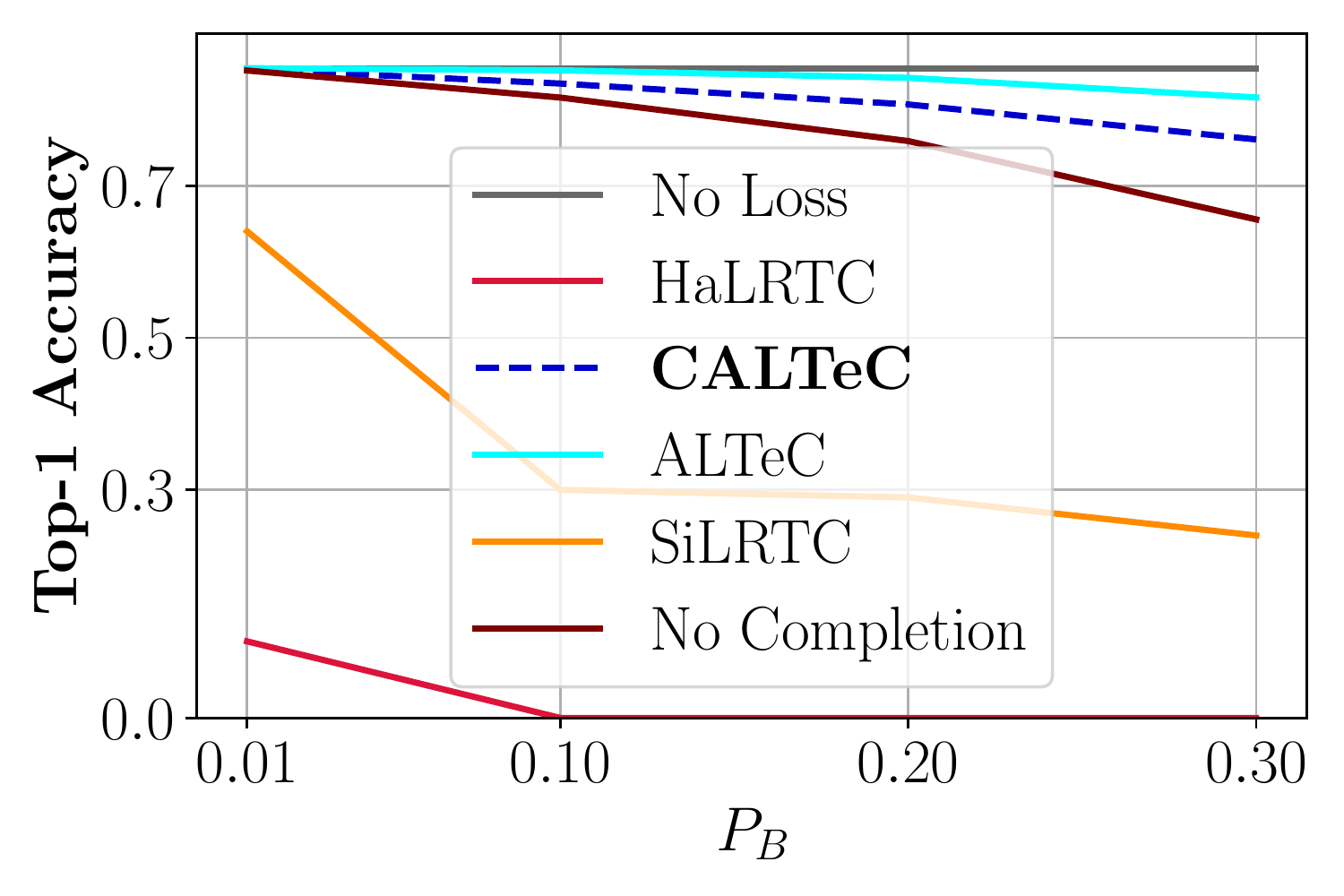}}
		\centerline{(d) Speed-matched \texttt{add\_3}.}
	\end{minipage} %
	\caption[Default and speed-matched results for two split layers]{Top-1 classification accuracy   averaged over 70 Monte Carlo simulation runs under various conditions. In (b), SiLRTC curve is behind the `No Completion' curve.}
	\label{fig:expt:defaultspeed}
\end{figure}

Top-1 accuracy at each $P_B \in \{0.01, 0.10, 0.20, 0.30\}$
is shown in Fig.~\ref{fig:expt:defaultspeed}. The value for each $P_B$ represents the average over 70 Monte Carlo runs (10 for each $L_B$). The figure shows two sets of results for each split point: default settings and speed-matched. Default settings correspond to running SiLRTC and HaLRTC for 50 iterations, which was observed in~\cite{Bragile2020} to be sufficient for their convergence.
ALTeC and CALTeC are non-iterative, and generally much faster, as seen in Table~\ref{tab:methods:speed}. Speed-matched results correspond to the case where each method is allowed to run for the amount of time that it takes CALTeC to complete its work. Iterative methods are allowed to complete the iteration in which their execution time exceeds that of CALTeC, i.e., they are allowed to run a bit longer. 
The entries for ALTeC do not include the several hours needed for pre-training, which is performed before operation of the system. ALTeC is faster than CALTeC, so it is able to complete its work in that amount of time. However, SiLRTC and HaLRTC are only able to complete one iteration on \texttt{add\_1} tensors and two iterations on \texttt{add\_3} tensors, due to the time-consuming singular value decomposition of unfolded tensors employed in these methods.  

Looking at default-settings results in Fig.~\ref{fig:expt:defaultspeed}(a) and (c), we see that all four tensor completion methods bring improvement compared to `No Completion,' except SiLRTC in Fig.~\ref{fig:expt:defaultspeed}(c). Among the four completion methods, SiLRTC has the worst performance. CALTeC and HaLRTC offer very similar, almost identical, performance. ALTeC is slightly worse than CALTeC and HaLRTC on \texttt{add\_1} tensors in Fig.~\ref{fig:expt:defaultspeed}(a), and slightly better on \texttt{add\_3} tensors in Fig.~\ref{fig:expt:defaultspeed}(c). However, keep in mind that ALTeC requires pre-training, whereas the other three methods do not.

In speed-matched results in Fig.~\ref{fig:expt:defaultspeed}(b) and (d), the advantage of CALTeC and ALTeC over SiLRTC and HaLRTC is obvious. CALTeC and ALTeC are non-iterative, and are able to recover missing data much faster than SiLRTC and HaLRTC. Especially HaLRTC needs around 10-15 iterations to start producing reasonable output; after just 1-2 iterations, the ``recovered'' data is of such poor quality that the resulting tensor virtually never produces the correct classification, so the Top-1 accuracy is near zero. The first 1-2 iterations of SiLRTC give better results, but still noticeably worse than CALTeC and ALTeC. Since inference latency is an important factor in collaborative intelligence, CALTeC and ALTeC have a clear advantage here over the other two methods. In addition, CALTeC does not require pre-training, so it combines the best properties of ALTeC (speed and accuracy) and those of SiLRTC/HaLRTC (no need for pre-training).

\begin{table}[t]
    \caption{Average execution times (in milliseconds) per test image for tensor completion methods used in the experiments. Values for SiLRTC and HaLRTC correspond to the time taken per iteration.}
    \vspace{4pt}
    \label{tab:methods:speed}
    \centering
    \begin{tabular}{l|r|r}
         Method & \texttt{add\_1} & \texttt{add\_3} \\
         \hline
         SiLRTC~\cite{liu2012tensor} (per iteration)& $228.1~\mathrm{ms}$ & $122.1~\mathrm{ms}$ \\
         HaLRTC~\cite{liu2012tensor} (per iteration) & $242.6~\mathrm{ms}$ & $128.2~\mathrm{ms}$ \\
         ALTeC~\cite{Bragile2020} & $30.5~\mathrm{ms}$ & $102.0~\mathrm{ms}$ \\
         CALTeC & $77.5~\mathrm{ms}$ & $186.8~\mathrm{ms}$
    \end{tabular}
\end{table}

\begin{figure}[!t]
	\centering
		\begin{minipage}[b]{0.31\linewidth}
		\centering
		\centerline{\includegraphics[width=2.5cm]{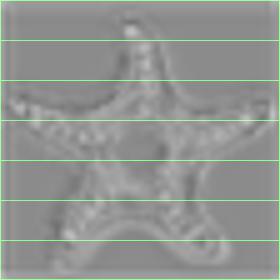}}
		\centerline{(a) Original}
	\end{minipage}
\begin{minipage}[b]{0.31\linewidth}
	\centering
	\centerline{\includegraphics[width=2.5cm]{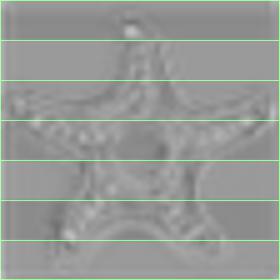}}
	\centerline{(b) ALTeC}
\end{minipage}
\begin{minipage}[b]{0.31\linewidth}
	\centering
	\centerline{\includegraphics[width=2.5cm]{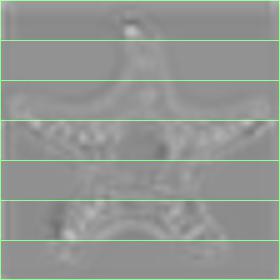}}
	\centerline{(c) CALTeC}
\end{minipage}
\hfill
\vspace{6pt}
	\centering
		\begin{minipage}[b]{0.31\linewidth}
		\centering
		\centerline{\includegraphics[width=2.5cm]{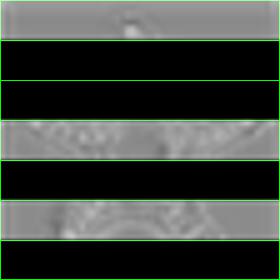}}
		\centerline{(d) Damaged}
	\end{minipage}
	\begin{minipage}[b]{.31\linewidth}
	\centering
\centerline{\includegraphics[width=2.5cm]{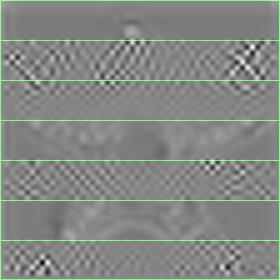}}
	\centerline{(e) SiLRTC 50}
    \end{minipage}%
	\begin{minipage}[b]{.31\linewidth}
		\centering
		\centerline{\includegraphics[width=2.5cm]{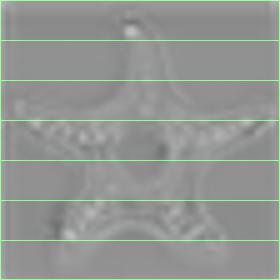}}
		\centerline{(f) HaLRTC 50} 
	\end{minipage} %
	\caption[Channel 1 visualization in a corrupted starfish image ResNet-18 tensor]{Channel 1 in a ResNet-18 \texttt{add\_1} tensor produced from a starfish image: (a) original; (b) damaged;  repaired with (c) CALTeC, (d) ALTeC, (e) SiLRTC and (f) HaLRTC. Images were mapped to grayscale and scaled up using bicubic interpolation for enhanced visualization.}
	\label{fig:expt:repair:starfish}
\end{figure}

Fig.~\ref{fig:expt:repair:starfish} shows an example of a damaged \texttt{add\_1} layer tensor channel (specifically, channel 1 from Fig.~\ref{fig:tensorviz}) and how it is repaired by various methods. CALTeC, ALTeC, and HaLRTC do a fairly good job of recovering the missing data, with starfish outlines clearly visible in their outputs. On the other hand, SiLRTC performs poorly in this example.
ALTeC in Fig.~\ref{fig:expt:repair:starfish}(d) appears to be most successful in recovering the underlying texture here, but note that CALTeC yields  better overall Top-1 accuracy on \texttt{add\_1} layer tensors over the whole test set, as shown in Fig.~\ref{fig:expt:defaultspeed}(a).

\section{Conclusions}
\label{sec:conclusions}
We presented Content-Adaptive Linear Tensor Completion (CALTeC), a method for recovering missing feature tensor data in collaborative intelligence. CALTeC takes advantage of intra- and inter-channel similarity in feature tensors to find the best candidates in other channels to fill the gaps in feature tensors, and then applies an affine transformation estimated from the same channel to properly adjust recovered data. In doing so, CALTeC combines the best properties of methods previously used for this purpose: speed, accuracy, and no need for pre-training. Experiments demonstrated CALTeC's competitive performance compared to other methods. 


\bibliographystyle{IEEEbib}
\bibliography{strings,refs}

\begin{thebibliography}{10}

\bibitem{neurosurgeon}
Y.~Kang, J.~Hauswald, C.~Gao, A.~Rovinski, T.~N. Mudge, J.~Mars, and L.~Tang,
\newblock ``Neurosurgeon: Collaborative intelligence between the cloud and
  mobile edge,''
\newblock in {\em Proc. ASPLOS'17}, Apr. 2017.

\bibitem{jointdnn}
A.~E. Eshratifar, M.~S. Abrishami, and M.~Pedram,
\newblock ``{JointDNN}: An efficient training and inference engine for
  intelligent mobile cloud computing services,''
\newblock {\em IEEE Trans. Mobile Computing}, vol. 20, no. 2, pp. 565--576,
  Feb. 2021.

\bibitem{choi2018deep}
H.~Choi and I.~V. Baji{\'c},
\newblock ``Deep feature compression for collaborative object detection,''
\newblock in {\em Proc. IEEE ICIP'18}, Oct. 2018, pp. 3743--3747.

\bibitem{eshratifar2019bottlenet}
A.~E. {Eshratifar}, A.~{Esmaili}, and M.~{Pedram},
\newblock ``{BottleNet}: A deep learning architecture for intelligent mobile
  cloud computing services,''
\newblock in {\em Proc. IEEE/ACM Int. Symposium Low Power Electronics and
  Design (ISLPED)}, Jul. 2019.

\bibitem{Chen20}
Z.~Chen, K.~Fan, S.~Wang, L.~Duan, W.~Lin, and A.~C. Kot,
\newblock ``Toward intelligent sensing: Intermediate deep feature
  compression,''
\newblock {\em IEEE Trans. Image Processing}, vol. 29, pp. 2230--2243, 2020.

\bibitem{Duan2020VideoCF}
L.~Duan, J.~Liu, W.~Yang, T.~Huang, and W.~Gao,
\newblock ``Video coding for machines: A paradigm of collaborative compression
  and intelligent analytics,''
\newblock {\em IEEE Trans. Image Processing}, vol. 29, pp. 8680--8695, 2020.

\bibitem{cohen2020lightweight}
R.~A. Cohen, H.~Choi, and I.~V. Baji{\'c},
\newblock ``Lightweight compression of neural network feature tensors for
  collaborative intelligence,''
\newblock in {\em Proc. IEEE ICME'20}, Jul. 2020.

\bibitem{vcm_call_for_evidence}
ISO/IEC,
\newblock ``Call for evidence for video coding for machines,'' ISO/IEC JTC 1/SC
  29/WG 2, m55065, Oct. 2020.

\bibitem{JPEG-AI_use_cases}
J.~Ascenso,
\newblock ``{JPEG AI} use cases and requirements,'' ISO/IEC JTC 1/SC29/WG1
  M90014, Jan. 2021.

\bibitem{choi_neural_2019}
K.~Choi, K.~Tatwawadi, A.~Grover, T.~Weissman, and S.~Ermon,
\newblock ``Neural joint source-channel coding,''
\newblock in {\em Proc. {ICML}}, Jun. 2019, pp. 1182--1192.

\bibitem{BottleNet++}
J.~{Shao} and J.~{Zhang},
\newblock ``Bottlenet++: An end-to-end approach for feature compression in
  device-edge co-inference systems,''
\newblock in {\em Proc. IEEE ICC Workshops}, Jun. 2020, pp. 1--6.

\bibitem{unnibhavi2018dfts}
H.~Unnibhavi, H.~Choi, S.~R. Alvar, and I.~V. Baji{\'c},
\newblock ``{DFTS: Deep} feature transmission simulator,'' \textit{IEEE
  MMSP'18} demo, 2018,
\newblock \url{https://github.com/SFU-Multimedia-Lab/DFTS}.

\bibitem{Bragile2020}
L.~{Bragilevsky} and I.~V. Baji{\'c},
\newblock ``Tensor completion methods for collaborative intelligence,''
\newblock {\em IEEE Access}, vol. 8, pp. 41162--41174, 2020.

\bibitem{liu2012tensor}
J.~Liu, P.~Musialski, P.~Wonka, and J.~Ye,
\newblock ``Tensor completion for estimating missing values in visual data,''
\newblock {\em IEEE Trans. Pattern Analysis and Machine Intelligence}, vol. 35,
  no. 1, pp. 208--220, Jan. 2013.

\bibitem{ResNet}
K.~{He}, X.~{Zhang}, S.~{Ren}, and J.~{Sun},
\newblock ``Deep residual learning for image recognition,''
\newblock in {\em Proc. IEEE CVPR'16}, Jun. 2016, pp. 770--778.

\bibitem{video2002}
Y.~Wang, J.~Ostermann, and Y.-Q. Zhang,
\newblock {\em Video processing and communications},
\newblock Prentice Hall, 2002.

\bibitem{5755057}
G.~{Hasslinger} and O.~{Hohlfeld},
\newblock ``The {Gilbert-Elliott} model for packet loss in real time services
  on the {Internet},''
\newblock in {\em Proc. 14th GI/ITG Conference - Measurement, Modelling and
  Evaluation of Computer and Communication Systems}, 2008, pp. 1--15.

\bibitem{imagenet2015}
O.~Russakovsky et~al.,
\newblock ``{ImageNet} large scale visual recognition challenge,''
\newblock {\em Int. J. Comput. Vision}, vol. 115, no. 3, pp. 211--252, Dec.
  2015.

\end{thebibliography}

\end{document}